# The nature of the electron

**Qiu-Hong Hu**

*Department of Physics, Gothenburg University, SE-412 96 Göteborg, Sweden, and LightLab Sweden AB, Smedjegatan 6, SE-131 34 Nacka, Sweden*

Through investigating history, evolution of the concept, and development in the theories of electrons, I am convinced that what was missing in our understanding of the electron is a structure, into which all attributes of the electron could be incorporated in a self-consistent way. It is hereby postulated that the topological structure of the electron is a closed two-turn helix (a so-called Hubius Helix) that is generated by circulatory motion of a mass-less particle at the speed of light. A formulation is presented to describe an isolated electron at rest and at high speed. It is shown that the formulation is capable of incorporating most (if not all) attributes of the electron, including spin, magnetic moment, fine structure constant $\alpha$, anomalous magnetic moment $\frac{1}{2}(\frac{\alpha}{\pi})$, and charge quantization into one concrete description of the Hubius Helix. The equations for the description emerge accordingly. Implications elicited by the postulate are elaborated. Inadequacy of the formulation is discussed.

**Resumé:**

En passant en revue le développement des théories de l'électron ainsi que l'histoire et l'évolution du concept lui-même, j'ai acquis la conviction que ce qui a manqué à notre compréhension de l'électron est une structure qui rassemblerait tous ses attributs de façon cohérente. Il est postulé ici que la structure topologique de l'électron est une doule hélice fermée (appelée Hubius helix en anglais) qui est crée par le mouvement circulaire d'une particule sans masse à la vitesse de la lumière. Une formulation est présentée pour décrire un électron isolé au repos et à grande vitesse. Il est démontré que cette formulation est capable d'incorporer en une description conctréte de la double hélice de Hubius la plupart (si ce n'est la totalité) des attributs de l'électron, y compris le spin, le moment magnétique, la constante de structure fine α, le moment magnétique anormal $\frac{1}{2}(\frac{\alpha}{\pi})$, et la quantization de la charge. Les équations de la description en découlent. Lesimplications du postulat sont développées et les insuffisances de la formulation sont discutées.

E-mail: qiuhong@physics.gu.se

# 1. CONCEPTS, HISTORY AND THEORIES

The electron as one of the elementary particles is perhaps as old as the universe, if one associates the hierarchical structure of the constituents of matter with the sequential events of creation and evolution of matter. The more elementary the building block of the constituent of matter, the earlier it should be created. However, as far as the origin of the word is concerned, "electron" was coined in 1894 by J. Stoney to denote a definite quantity of electricity. [1] The discovery of the electron as an isolated entity in the form of the cathode ray particles was attributed to J. J. Thomson by whom they were referred to as corpuscles and their charge as the electron. [2] In later usage, however, these particles themselves came to be called electrons. While J. Stoney emphasized the atomic nature of electricity, J. J. Thomson's discovery and the subsequent work in his school established firmly that the electron is an elementary constituent of an atom and therefore of matter. Since then, the electrons have provided mankind with an ever-richer content of not merely physics, but science, technology, and our daily life at large.

The history of the electron can be viewed in two ways, either as consisting of a sequence of epochal events in which new attributes of the electron were revealed or as consisting of time periods marked by decisive progress in the conceptual understanding of the electron. Insofar as the isolated entity is concerned, the epochal events are:

- Discovery of the electron in the 1890's, based on several independent experiments a negative charge *e*, and mass *m* were assigned to the electron,

- Recognition and verification of the increase of effective mass with velocity in 1905 and 1906,

- Assignment of the Compton wavelength $\lambda_c = \dfrac{h}{mc}$ to the electron from the scattering of X-rays by free electrons in 1923,

- Hypothesis of wave properties of the electron by de Broglie in 1924, $\lambda = \frac{h}{mv}$, which was verified by C. J. Davisson and L. H. Germer's experiment of electron diffraction by crystals in 1927, and independently by G. P. Thomson in 1928,
- Postulate of the concept of the spinning electron in 1925 and 1926 by G. Uhlenbeck and S. Goudsmit, in which the spin angular momentum of $\frac{1}{2}\hbar$ and magnetic moment of $\frac{e\hbar}{2mc}$ were assigned to the electron,
- The discovery of the Dirac equation in 1928,
- Derivation of Zitterbewegung of the electron in 1930, by analyzing the Dirac equation in the Heisenberg representation, E. Schrödinger arrived at the fact that there must be an oscillatory motion associated with a free Dirac electron, which he named Zitterbewegung [†],
- Prediction of the anti-electron (positron) by P. Dirac in 1931, which was discovered by C. Anderson in 1932,
- Existence of the anomalous magnetic moment of the electron, which was put into question by G. Breit in 1947 and experimentally determined by P. Kusch and H. M. Foley in 1948.

It was through these events that the concept of the electron was established and its attributes were revealed. The event-sequence view puts emphasis on the physical content of the concept of the electron, as a complement the time-period view provides a scenario of what man has achieved in understanding of the concept.

According to the time-period view, the history of the electron can be divided into four periods. The first period appears to start with a merging of knowledge from three then (1880's) totally unrelated disciplines of experimental science, electrolysis, cathode rays and radioactivity. It ended up with the discovery of the electron that was of particle nature and its identification as an elementary constituent of the atom. The second period, up to the discovery of the wave properties of the electron, is the history of the electron as a relativistic particle according to Abraham, Lorentz, Einstein and Poincaré. The third period was a period of striving to understand and finally accept the wave and spin properties of the electron through formulating wave mechanics and relativistic quantum theory of the electron. This period finished with the establishment of quantum electrodynamics (QED) that left us with a renormalization picture of the electron. The post QED era of the electron's history was conjectured as the fourth period, "a nonperturbative internal structure of the electron". [3]

Developments in the theories of the electron can be regarded as being evolved from the classical theory, through the relativity theory, the non-relativistic quantum theory, and relativistic quantum theory to quantum electrodynamics and to refining and extending the renormalization theory of the electron in the last 50 years. For reviews of the developments in the theory of the electron prior to QED readers are referred to Pais and Weisskopf. [4, 5] For reviews of the developments in QED and quantum field theories see R. P. Feynman and W. Heitler respectively. [6, 7] Various aspects of the latest development in electron theory and QED were summarized in a recent volume of the NATO ASI series. [8]

## 2. WHAT WAS MISSING?

In a little over one hundred years of the electron's history, a tremendous progress has been achieved in details and applications of the electron theories.[9] While electrons are useful in the understanding of diverse phenomena, we have never really understood the electron itself. I am convinced that what was missing was the very nature of the electron—a structure, into which all attributes of the electron could be incorporated in a self-consistent way.

The nature of the electron has been an enigma confronting physicists of past generations as well as contemporaries.[10, 11] The whole issue involved was typified by the late Asim Barut's thought provoking questions (see reference 3): what is really an electron? What is the structure that gives its spin? What is the structure that can appear under probing with electromagnetic fields as a point charge, yet as far as spin and wave properties are concerned exhibits a size of the order of the Compton wavelength? Why must there be a positron? What is mass? Why and how does the electron manifest wave properties? And what is the interaction between two electrons or between an electron and a positron at short distances? A question of curiosity to seek for an intuitive picture of the electron may turn out to have far-reaching consequences, for "Theoretical physics has made use, for a long time, of abstract representations. ... They are indeed very useful and even almost essential auxiliaries of reasoning. But one must never forget that the abstract presentations have no physical reality. Only the movement of elements localized in space, in the course of time, has physical reality".[12]

The abstraction, on the one hand, seeks to crystallize the logical relations inherent in the maze of material that is being studied, and to correlate the material in a systematic and ordered manner. Intuitive understanding, on the other hand, fosters a more immediate grasp of the objects one studies, a live *rapport* with them, which stresses the concrete meaning of their relations. [13] While the abstract presentation prevails, a live *rapport* of the electron has not emerged over the years.

"In speculating on the structure of these minute particles", Lorentz remarked as early as 1909, "we must not forget that there may be many possibilities not dreamt of at present; and perhaps, after all, we are wholly on a wrong track when we apply to the parts of an electron our ordinary notion of force". [14] Fermi wrote in 1932 in concluding his article on quantum theory of radiation, "In conclusion, we may therefore say that practically all the problems in radiation theory which do not involve the structure of the electron have their satisfactory explanation; while the problems connected with the internal properties of the electron are still very far from their solution". [15] In introducing his relativistic quantum theory of the electron in 1928, Dirac posed a question "why Nature should have chosen this particular model [an electron with a spin angular momentum of half a quantum and a magnetic moment of one Bohr magneton] for the electron instead of being satisfied with the point-charge". [16] As triumphs of the theory, spin $\frac{1}{2}\hbar$ and magnetic moment $\frac{e\hbar}{2mc}$ of the electron were contained in Dirac's mathematical construction, but left unexplained. [17] The following twenty years witnessed fights against infinities appearing in the interaction of the electron with electromagnetic fields and finally the establishment of renormalized QED. [18, 19] In spite of its tremendous success in calculations, QED cannot be considered as a satisfactory solution to the problem of the electron. [20] Being dissatisfied with the renormalization treatment of the infinities, Dirac pinpointed the trouble of QED being working from a wrong classical theory, and put forward a new classical theory of the electron as a basis for a passage to the quantum theory. [20-22] Unfortunately he did not proceed to provide a clue as to how to bring in charge quantization, spin and Fermi statistics for the electrons in making the passage. Viewing the discovery of the muon as new evidence for the finite size of the electron and believing "The spin angular momentum of a particle should be pictured as due to some internal motion of the particle", [23] Dirac returned to the classical theory of the electron by modelling an electron as an extensible object. [24] A particular difficulty in developing his idea is to bring in the correct spin. Attempts of other people have also been made over the years aiming at resolving the problems associated with the structure of the electron starting from self-energy, size and structure, spin, and Zitterbewegung, [3, 11, 25-28] to list just a few.

It is understood that our existing knowledge of the electron imposes immense constraints on, and yet provides hardly any hint to a self-consistent formulation of the structure. In this article, a topological structure of the electron is postulated as a working hypothesis. As consequence and validity test of the postulate, a formulation is presented to describe an isolated electron at rest and moving at high speed. It is shown that the formulation is able to incorporate most, if not all, attributes of the electron, including spin $\frac{1}{2}\hbar$, magnetic moment $\frac{e\hbar}{2mc}$, fine structure constant $\alpha$, anomalous magnetic moment $\frac{1}{2}(\frac{\alpha}{\pi})$ and charge quantization into one concrete description of the topological structure—one particular space curve. The equations for the topological structure emerge accordingly. Implications elicited by the postulate are elaborated. Inadequacy of the formulation is discussed.

## 3. THE POSTULATE

The topological structure of the electron is a closed two-turn helix (a so-called Hubius Helix) that is generated by circulatory motion of a mass-less particle at the speed of light.

The Hubius Helix can be expressed analytically by

$$\begin{cases} x = (R + r\sin\frac{\theta}{2})\cos\theta \\ y = (R + r\sin\frac{\theta}{2})\sin\theta, \text{ Where } R, r \in \mathbf{R}_e, \ R > r > 0; \\ z = r\cos\frac{\theta}{2} \end{cases} \quad (1)$$

and be depicted graphically in Fig. 1.

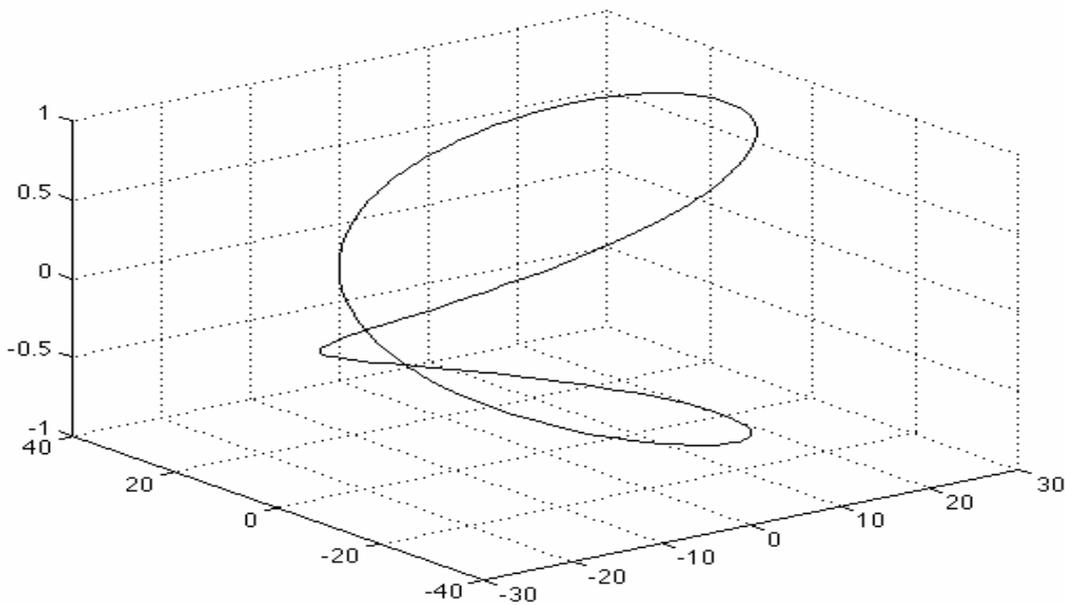

Fig. 1 Graphical representation of the Hubius Helix created from the parameterized equations (1) using Matlab.

# 4. THE FORMULATION

## 4.1. Point of Departure: $mc^2 = h\nu$

In 1905 Einstein proposed two basic equations, $E = mc^2$ in the special theory of relativity for particles of the rest mass *m*, and $E = h\nu$ in the theory for photoelectric effect for photons of the frequency $\nu$. About twenty years later L. de Broglie asserted that $E = h\nu$ applied to particles of every sort. Then the two equations yield the relation $mc^2 = h\nu$, which strongly suggests that some kind of oscillatory process be associated with matter. L. de Broglie himself always believed that the relation described real physical internal oscillations. To L. de Broglie, the relation defines the frequency as a function of mass or vice versa. From this relation L. de Broglie derived $\lambda = \dfrac{h}{mV / \sqrt{1 - \dfrac{v^2}{c^2}}} = \dfrac{h}{p}$ the "de Broglie wavelength" for a particle moving at *V* to a fixed observer. [29]

## 4.2. Quantization on Hubius Helix

There have been several alternative ways to establish quantization conditions in the development of the quantum mechanics, among others, Bohr-Sommerfeld-Wilson's, de Broglie's and Schrödinger's quantization conditions. Bohr-Sommerfeld-Wilson's quantization condition introduced the action integral familiar in classical mechanics into quantum theory, acted as an intermediate step in the discovery of nonrelativistic quantum mechanics, and was superseded by the latter. It may still give correct results in a limited number of problems in the cases of a single variable, and when the separation of variables is possible in the case of many degrees of freedom, Chapter 10 in [30] and [31], where the proof of equivalence of these quantization conditions are given as well. The quantization on the Hubius Helix concerns a variable of one degree of freedom, thus the Bohr-Sommerfeld-Wilson's quantization condition is applicable. However, in view of the great difficulties and serious shortcomings, the construction of quantization condition on the Hubius Helix will not follow this alternative. Schrödinger's quantization condition is based on a variation principle, as well as the analogy of mechanics and optics. The first derivation was based on a "heuristic" argument resembling the variation of average energy of the system, and later was thought by him as "incomprehensible". In the second derivation, he added the analogy of geometric optics and wave optics and of particle and wave mechanics to put his equation on a secure ground by the variation of the Hamiltonian operator of a quantum system, Chapter 11 in [30] and [32]. Schrödinger's alternative treated the electron in the Coulomb field as a dynamic system. Through finding the quantum Hamiltonian of the system and applying variation principle to the action associated with the Hamiltonian, the problem of quantization is converted to the eigenvalue problem of differential equation, and the particle was replaced by the wave. The intention of the present work is not to treat the electron as a dynamic system, but as a topological/geometrical object or a spatial temporal structure, into which its physical attributes are then to be incorporated. From this point of view, the construction of quantization condition on the Hubius Helix is thought to be in line with the geometrical argument put forward by de Broglie.

De Broglie's quantization condition emerged in the end of the proof that the Maupertuis' principle of action for a particle moving a along a path is identical with the Fermat's principle for the waves accompanying the particle:

$$\int_C \frac{dl}{\lambda} = \int_C \frac{1}{h} p\, dl = Integer \qquad (2)$$

In order to arrive at the proof, de Broglie verified that the group of the waves associated with the particle has the same velocity as the particle, and that the paths of the particles are identical with the rays of the waves. [29, 30]

Equation (2) can be rewritten separately as $\oint_C dl = n\lambda,\ and \oint_C pdl = nh$ on a closed path. The former equation implies that the wave along the path must be integer multiples of a fundamental wavelength, the latter implies that the action along the path must be integer multiples of the action quantum, and the connection of the two equations bestows a geometrical meaning to the Bohr-Sommerfeld-Wilon's quantization condition.

Applying de Broglie's quantization condition on the Hubius Helix, the quantization on the Hubius Helix emerges

$$\oint_l \frac{1}{\lambda} dl = \int_0^{4\pi} \frac{1}{\lambda} r d\theta = n \tag{3a}$$

Or $\quad 4\pi r = \lambda \tag{3b}$

The $r$ is the radius of the Hubius Helix. Put $n=1$ for the fundamental wave.

### 4.3. Compton Wavelength, Mass and Radius of the Electron

Equation (3b) implies a geometric argument that the Hubius Helix must contain a whole wave of $4\pi$ periodicity, an argument akin to the original standing wave argument of de Broglie. To put the argument into the context of the electron, one starts with $mc^2 = h\nu$, combine $\lambda = \frac{V}{\nu}$, where $V$ is the velocity of the wave ($V = c$, according to the postulate), and arrives at $r = \frac{\hbar}{2mc} = \frac{\lambda_c}{4\pi}$ through algebraic manipulation. It is noteworthy that the unspecified wavelength introduced in the quantization condition turns out to be the Compton wavelength of the electron. The radius of the Hubius Helix, denoted $r_e$ hereafter, is defined as the radius of the electron.

The expression $r = \dfrac{\hbar}{2mc}$ has been assumed *a priori* or derived several times in the past. [25, 33, 34] Unfortunately, it was not accepted unequivocally as the radius of the electron. The facts that it can be derived from the quantization condition and that it does not contradict other attributes of the electron seem to justify its acceptance as the radius of the electron (the second fact will be shown in the proceeding subsections).

**4.4. The Spin**

An orbital motion gives rise to an angular momentum by definition in classical mechanics. Following the notion of circular motion, the spin *s*, is identified as the orbital angular momentum of the mass-less particle performing circulatory motion, thus: $s = mvr_e = mc\dfrac{\hbar}{2mc} = \dfrac{1}{2}\hbar$.

**4.5. The Magnetic Moment**

A current loop gives rise to a magnetic moment, or generally all magnetic fields are due to the motion of electric charges, according to Ampere's hypothesis. [6, 35] For a planar current loop, the magnitude of the magnetic moment is $\mu = \dfrac{nIA}{c}$, where *n* is the number of turns of the loop, *I* is the current, and *A* is the area of the loop. [36] By applying Ampere's hypothesis to the postulated structure, and taking into account the fact that the two-turns suggests that *n*=2, one gets $\mu_e = \dfrac{nIA}{c} = 2\dfrac{ec}{2\pi c r_e} \times \pi r_e^2 = er_e = \dfrac{e\hbar}{2mc}$.

The spin and the magnetic moment are intricately connected in the understanding of the electron. [37] The original spinning electron hypothesis was deduced phenomenologically to account for the anomalous Zeeman Effect and is in agreement with experiment. [38, 39] It states "An electron behaves as a charged particle which is rotating around an axis in such a way, that it possesses an angular momentum of constant absolute magnitude $\frac{1}{2}\hbar$ around that axis and that it possesses at the same time a magnetic moment $\frac{e\hbar}{2mc}$ which has a direction opposite to the direction of the angular momentum". [39] By convention the direction of an electric current is opposite to that of the flow of the negative charge. According to this convention, the spin and the magnetic moment derived in the formulation are in the opposite direction, which is consistent with the spinning electron hypothesis. Furthermore, the formulation, an extension of the hypothesis, provides nothing but "such a way" that was left out in the spinning electron hypothesis.

## 4.6. The Positron

Dirac predicted that in addition to electrons there must be another particle, called positron that is necessarily related to the electron. [40] The positron is also called the antiparticle of the electron. The properties of a particle and its antiparticle obey certain rules of correspondence: they possess an equal amount of energy or mass and the opposite charge. More importantly, when a particle and its antiparticle come together, they can annihilate each other and liberate their entire mass in the form of radiation energy, say $\gamma$-rays. In the case of an electron and a positron, the annihilation proceeds as $e^- + e^+ = \gamma + \gamma'$. Conversely, an electron and a positron can be pair-produced following the reverse reaction. The rules of correspondence demand that an electron and a positron be treated on the same footing, i.e. a theory applied to the electron should apply equally to the positron. [41]

If I am to formulate the positron in terms of Hubius Helix satisfying the rules of correspondence between an electron and a positron, I am bound to identify one feature of the Hubius Helix to account for the equality in the mass of the electron and positron, another feature to account for the disparity in the charge of the electron and positron, and yet a mechanism for creation and annihilation of mass and charge to account for the creation and annihilation of an electron-positron pair.

According to the relation of the mass of the electron to the radius of curvature of the Hubius Helix $r_e = \frac{\hbar}{2mc}$, the fact that an electron and a positron are of the same mass can be stated equivalently as that they are of the same radius of curvature, i.e. if $r_{e^-} = \frac{\hbar}{2mc}$, and $r_{e^+} = \frac{\hbar}{2mc}$; then $r_{e^-} = r_{e^+}$. Given the radius of curvature, winding determines a Hubius Helix right-handed or left-handed: the right-hand screw winding results in a right-handed Hubius Helix, whereas the left-hand screw winding results in a left-handed Hubius Helix. A right-handed Hubius Helix cannot be congruent to a left-handed Hubius Helix, except by a mirror reflection. Winding does not only determine the parity of a Hubius Helix, it also results in a twist. In spite of the fact that the magnitude of the twist depends on parameters that specify the details of the winding, the sign of a twist can be defined positive/negative with respect to the handedness of winding. Once the relation of the parity of the Hubius Helix to the sign of twist is defined, a change in parity will necessarily lead to a change in the sign of the twist, and vise versa. Thus the disparity in the charge of the electron and positron can be connected to the parity or the sign of the twist of the Hubius Helix. This connection is qualitative and incomplete in character, and its completeness is grounded on a quantitative relation between the charge and the magnitude of the twist, a question concerning the origin of the charge and charge quantization (see subsection 4.10).

Following the curvature statement and the connection of the sign of the charge to the parity and the sign of the twist, the rules of correspondence of an electron to a positron can be formulated as such that an electron and a positron are two Hubius Helices of the same radius of curvature but opposite twist and parity. The annihilation is hence to be understood as unwinding of the two Hubius Helices of the same radius of curvature and the opposite parity and twist, whose consequence is the creation of two photons with the opposite polarization. Conversely, the pair production is the confinement of two photons in the form of Hubius Helix of the same radius of curvature and the opposite parity or twist. These geometrical rules of correspondence possess a beautiful feature in that an electron and a positron—two physical entities are unified to a single geometric entity.

## 4.7. Hubius Helix and Dirac Electron

A Dirac electron is the electron described by Dirac's relativistic quantum theory of the electron. The Dirac theory of the electron or the Dirac equation is an ingenious amalgamation of the special theory of relativity and quantum mechanics on the electron to account for the so-called duplexity phenomena in atomic spectra without arbitrary assumption. [16] In order for the theory to be compatible with both quantum mechanics and the special theory of relativity, the requirements of the general schemes of both must be fulfilled. For quantum mechanics the general scheme requires that only first order derivatives with respect to time appear. On the other hand, the general scheme of relativity requires that the time and space co-ordinates be treated on the same footing. In satisfying these requirements, the introduction of Dirac's matrices **α** and **β** in the Hamiltonian $H = c\mathbf{\alpha} \cdot p + mc^2 \mathbf{\beta}$, brought about "four times as many solutions as the non-relativity wave equation and twice as many as the previous relativity wave equation", which implies the existence of a new particle of positive elementary charge associated with the electron. [16] (It became known later that two solutions belong to an electron in its two spin eigenstates, and two belong to a positron in its two spin eigenstates.). [42] When the motion of the electron in a central field was investigated, an extra contribution of $\frac{1}{2}\hbar$ was obtained in the total angular momentum. This extra contribution was interpreted as the spin angular momentum of the electron. The magnetic moment terms were obtained in a similar way, when scalar and vector potentials were included in the Hamiltonian. Thus, the Dirac theory of the electron contains many new features, which were absent in previous theories. These new features in turn bestow new features on the Dirac electron. The Dirac electron is associated with an anti-electron of positive charge in such a way that the two eigenstates of the spin of an electron and those of the anti-electron are to be described as four components of one entity represented by a column matrix of four elements. A Dirac electron carries with it a spin $\frac{1}{2}\hbar$ and magnetic moment $\frac{e\hbar}{2mc}$.

The postulate through the formulation of subsections 4.4, 4.5 and 4.6 provides a description of the electron consistent with the Dirac theory, by providing the electron with the same features as those of the Dirac electron: a spin $\frac{1}{2}\hbar$ and magnetic moment $\frac{e\hbar}{2mc}$, and the same electron-positron association. In the Dirac theory, these features are revealed through algebraic manipulation, whereas in the present formulation, these features appear as natural consequences of the topological and geometrical features of the Hubius Helix. The consistency indicates that the postulate contain a great deal of truth. Furthermore, the consistency, if reflecting the inner consistency of physical laws, should no doubt lead to a profound connection between the Dirac theory of the electron and the Hubius Helix. It turns out that such a connection is embedded in another feature of the Dirac theory, the spinor.

The name spinor was first introduced by Ehrenfest. [43] While the general mathematical form of spinors was discovered by E. Cartan, [44] it was the re-invention by Dirac that provided the first indications of its fundamental importance to physics, the four wave functions in the Dirac equation being only the components of a spinor. However, it is believed that we have still not seen the full significance of spinors in the basic structure of physical laws. [45] Based on my recognition that a spinor changes sign when it completes a $2\pi$ rotation and returns to its original position after $4\pi$ rotation is exactly the property that a Hubius Helix possesses, [46] and that the components of a spinor are the same in number as the variants of a Hubius Helix (each handedness corresponding to two spin directions); it is hereby conjectured that the structure of a spinor is the Hubius Helix. Consequently, the Hubius Helix is the geometrical manifestation of the Dirac electron.

**4.8. Circular Motion and Zitterbewegung**

In examining the position and velocity operators associated with the Dirac equation in the Heisenberg representation, E. Schrödinger found that the eigenvalue of the three space components of velocity operator $c\vec{\alpha}$ could only take $\pm c$ (the same conclusion was reached earlier by G. Breit). And that the position operator consists of two parts, $x_k = \tilde{x} + \xi_k$ in which $\tilde{x} = a_k + c^2 H^{-1} p_k t$ and $\xi_k = -\frac{c\hbar}{2i} \eta_k^o H^{-1} \exp(-i\frac{2Ht}{\hbar})$. The $k$ denotes the $k$th component of the operators. The rectilinear part he interpreted as the position operator of the center-of-mass, whereas the resultant part he interpreted as the position operator of the charge in the rest frame. He thus concluded that the motion of a free Dirac electron consists of a rapidly oscillatory component—*Zitterbewegung*, superimposed on the average motion. [33]

The Zitterbewegung, as a concept and consequence derived from the Dirac theory contains the following physical contents: the instantaneous velocity of the charge being always $c$, the linear dimension (radius to be exact) of the charge cloud and the characteristic amplitude of the Zitterbewegung being $r = \frac{\hbar}{2mc}$, and the characteristic angular frequency of the Zitterbewegung being $\omega_z = \frac{2mc^2}{\hbar}$. The present author is aware of the point of view that when the Dirac equation is interpreted as a many-particle equation or a field equation, as the perception angle moved from describing the motion of single particle to quantizing a field, as the picture of single particle disappears, so does the manifestation of the Zitterbewegung. This point of view denies the existence of the Zitterbewegung. However, as Dirac argued "One must believe this consequence of the theory, since other consequences of the theory which are inseparably bound up with this one, such as the law of scattering of light by an electron, are confirmed by experiment". [47] In spite of Dirac's argument, little attention has been given over the years to the physics of the Zitterbewegung.

Among the few exceptional cases it is shown that Zitterbewegung may be looked upon as a circular motion of the charge on the plane perpendicular to the direction of the electron spin and as arising from the superposition of the positive and negative frequency components of the spinor. A positive or negative component alone does not produce a measurable effect of Zitterbewegung.[48] It was suggested that one distinguish two reference frames, the center-of-charge and the center-of-mass.[49-51] When the Zitterbewegung is identified with the motion of the charge relative to the center-of-mass rest frame, the Hamiltonian becomes $H_r = mc^2 \beta$, and consequently position, velocity and momentum operators of the charge are identified with, $\vec{Q} = -i\frac{\hbar}{2mc}\beta\vec{\alpha}$, $\frac{d\vec{\xi}}{dt} = c\vec{\alpha}$ $\vec{P}_{charge}(t) = mc\vec{\alpha}(t)$, respectively.[50, 51] The identification clarifies the ambiguity in describing the Zitterbewegung, due to the difference in definition of the position operators.[52] Based on these results,[48-51] the pro Zitterbewegung view seems to make the case that Zitterbewegung is a self-consistent description of the dynamics of a free Dirac electron.

The fact that both the Zitterbewegung and the postulate require a mass-less particle performing circular motion at the velocity of *c*, lead to the same radius of the electron and therefore to the same angular frequency indicates that the dynamics entailed in the postulate is consistent with the dynamics of the Zitterbewegung.

Questions arise immediately, however, when the Zitterbewegung frequency $\omega_z = \frac{2mc^2}{\hbar}$ and the de Broglie frequency $\omega_d = \frac{mc^2}{\hbar}$ are put in juxtaposition. Can the de Broglie frequency be identified with the Zitterbewegung frequency as found in the literature?[26, 27] If not, which is the characteristic frequency of the electron, if either? Or else, does the electron have two characteristic frequencies?

The consistency between the Zitterbewegung and the postulate demands the acceptance of the Zitterbewegung frequency as a characteristic frequency of the electron. On the other hand, the expression for the de Broglie frequency—the wave-particle duality is physically valid in the broadest sense known. For these two reasons, one is compelled to accept a consequence that one and the same mass of the electron is associated with two frequencies. If such a consequence appears perplexing and irremediable within the context of Zitterbewegung and wave-particle duality, it is exactly an outcome of the mathematical structure inherent in the postulate—the frequency condition for forming a closed two-turn helix, as exemplified by

$$\begin{cases} \begin{cases} x = R\cos\frac{1}{2}\omega t \\ y = R\sin\frac{1}{2}\omega t \end{cases} \\ \begin{cases} x = R\cos\omega t \\ y = R\sin\omega t \end{cases} \end{cases}$$

in an idealized two-dimensional case. To put the equations into the context of the electron, $R$ is identified with the radius $r_e$, $\omega$ with the Zitterbewegung frequency, and $\frac{\omega}{2}$ with the de Broglie frequency.

### 4.9. Fine Structure Constant and Anomalous Magnetic Moment

The constant $\alpha = \frac{e^2}{\hbar c} \approx \frac{1}{137}$ that appears in the formula denoting the energy of the bound electron in a hydrogen atom and gives the amount of fine structure splitting in the line spectrum is known as the fine structure constant after Sommerfeld. It is a fundamental constant in the physical theories concerning electromagnetic interactions. [53] The uniqueness of $\alpha$ lies in the fact that it is a combination of three fundamental physical constants, and it is dimensionless. In spite of its unique position and importance in the physical theories, no theory for $\alpha$ is yet available. [53, 54] It is maintained improbable that a theory for $\alpha$ can be found before the new features in the description of nature, due to a fundamental length $\lambda$ associated with elementary particles, have been clarified, features which at first have no connection at all with the question of the electronic charge. [55] The $\lambda$, as a natural consequence of conjoining quantum mechanics and the special theory of relativity according to Heisenberg, cannot be formed dimensionally from $\hbar$ and $c$ *should* instead be related to the mass of the particle in the form $m \propto \frac{\hbar}{c\lambda}$. And its relation to $r_0 = \frac{e^2}{mc^2}$, the classical radius of the electron *should* be analogous to that of $h$ to $\hbar$. [55] Being a dimensional formula, the expression is not expected to produce the exact value; one needs a numerical coefficient, say $k$ to produce equality. In other words, one needs a physical reason to provide the coefficient. Following Heisenberg's profession as a guideline, however, one can try to identify what is the fundamental length associated with the electron and to find a way leading to $\alpha$. One can identify $r_e = \frac{\hbar}{2mc}$ with $\lambda$, and get $k=1/2$. It seems that $r_e$ is the fundamental length Heisenberg anticipated, but it does not bring $\alpha$ into being. One can identify, $r_0$ with $\lambda$, and get $k = \frac{e^2}{\hbar c} = \alpha$. Combining the result of the two trials, a new expression for $\alpha$ is found $\alpha = \frac{r_0}{2r_e}$. It is indeed an association of fundamental length(s) with the elementary particle, the electron in this case. The new expression $\frac{r_0}{2r_e} = \frac{e^2}{\hbar c}$ can also be obtained by identifying the rest energy of the electron with the Coulomb energy of the classical electron $mc^2 = \frac{e^2}{r_0}$, and substituting $m$ with $\frac{\hbar}{2cr_e}$. The connotation of the first step is that the electron mass is electromagnetic in origin in a sense of classical electrodynamics, and the connotation of the second step is the acceptance of $r_e$ as the fundamental length. In spite of the connotations, the essence of the new expression for $\alpha$ remains the ratio of two lengths associated with the electron. It should be noted, however, that while $r_e$ represents the radius of the Hubius Helix, $r_0$ does not have a geometrical meaning associated with the Hubius Helix. It is not unreasonable to assert that if one can identify the association of $r_0$ and the Hubius Helix, one is in a good position to account for $\alpha$ geometrically. "It seems hopeless to attack this problem [to obtain a theory of electrons which fixes the value of $\frac{e^2}{\hbar c}$] from the physical point of view, as one has no clue to what new physical ideas are needed. However, one can be sure that the new theory must incorporate some very pretty mathematics, and by seeking this mathematics one can have some hope of solving the problem." [22] Following Dirac's line of reasoning one ought to seek for the right mathematical content of the Hubius Helix.

It is important to note that the formulation outlined so far has dealt implicitly in two-dimensions with the radius, the periodicity and the frequencies of the circulatory motion. In other words, it has dealt with the topological aspect of the Hubius Helix and its one homotopy of a constant radius in two dimensions. The natural extension of the formulation is therefore to examine the Hubius Helix in three-dimensions.

In doing so, I did not fail to perceive that the surface bounded by a Hubius Helix is in fact a Möbius strip [56]—a one-sided non-orientable surface discovered by J. B. Listing and A. F. Möbius independently in 1858, and that the edge of a Möbius strip is a Hubius Helix. Such a coupled double perception (CDP) or visual dichotomy manifests itself as ambiguous figures in the worlds of art and cognitive sciences. [57] Well-known examples are the Necker Cube, where the CDP is produced by *perspective reversal*; Rubbin's Reversible Goblet, where the CDP is produced by *figure-ground reversal*; and Hill's My Wife and My Mother-in-law, where the CDP is produced by *rival-schemata reversal*. It is worth emphasizing that the Möbius strip-Hubius Helix dichotomy is beyond these categories of reversal. It is *boundary-the enclosed reversal* that produces the CDP. In spite of the ambiguity in perception, in many instances, the contents of ambiguous figures can be stated unambiguously in terms of mathematics. As in the case of the Möbius strip-Hubius Helix dichotomy, the Möbius strip as a surface is most simply represented in terms of two real parameters $\rho$ and $\varphi$ as follows:

$$\begin{cases} x = (a - \rho \sin\frac{1}{2}\varphi)\cos\varphi \\ y = (a - \rho \sin\frac{1}{2}\varphi)\sin\varphi \\ z = \rho \cos\frac{1}{2}\varphi \end{cases}. \quad (58)$$

For a given constant $a$, $|\rho| < a$ determines the width of the Möbius strip in the concrete description of each homotopy. On the other hand, the Hubius Helix as a space curve can be concretely described by the same equations in terms of only one real parameter $\varphi$ with $a$ and $\rho$ as constants. It seems that the Möbius strip as a non-orientable surface and non-trivial fiber-bundles has caught considerable attention in mathematics and physics, [59] and that its edge as a space curve has been overlooked, if not totally ignored. [60]

Thus, by examining the Hubius Helix in three dimensions, the relation between the Hubius Helix and the Möbius strip was found, and a new length $\rho$ was introduced, which defines the spacing between the two turns of the Hubius Helix. In addition, it is observed that while different topological objects may share the same concrete description, the same topological object can have completely different concrete descriptions. It is therefore anticipated that one of the concrete descriptions of the Hubius Helix be able to further relate $r_e$ or $r_0$ to $\rho$ and thereby to incorporate $\alpha$ into the postulated structure.

In subsection 4.5, the magnetic moment of the electron was expressed as $\mu_e = er_e$ in the idealized two-dimensional case. In the three-dimensional case, the instantaneous radius of the circular motion is a periodic function of time or phase according to $r = \sqrt{x^2 + y^2 + z^2}$. Thus, the measured radius should be the mean value of the instantaneous radii over the interval of $4\pi$ according to $\bar{r}_e = \frac{1}{4\pi}\int_0^{4\pi} r\, d\theta$. The measured magnetic moment should in turn be the product of the charge and the mean radius in the form of $\mu_e = e\bar{r}_e$.

Consider $\begin{cases} x = (r_e + 2\rho\sin\phi)\cos\theta \\ y = (r_e + 2\rho\sin\phi)\sin\theta \\ z = \rho\cos\phi \end{cases}$,

where $\rho, r_e \in R_e$, $r = \sqrt{x^2 + y^2 + z^2}$, with $\begin{cases} \rho = 0, (0 \leq \theta \leq 2\pi, 0 \leq \phi \leq \pi) \\ \rho < r_e, (2\pi \leq \theta \leq 4\pi, \pi \leq \phi \leq 2\pi) \end{cases}$.

The domains are under the constraints: $\phi|_{\theta=0} = 0$, $\phi|_{\theta=2\pi} = \pi$, and $\phi|_{\theta=4\pi} = 2\pi$.

Then

$$\bar{r}_e = \frac{1}{4\pi}\int_0^{4\pi} \sqrt{x^2 + y^2 + z^2}\, d\theta$$
$$= \frac{1}{4\pi}\int_0^{2\pi} \sqrt{r_e^2\cos^2\theta + r_e^2\sin^2\theta + \rho^2\cos^2\phi}\, d\theta + \frac{1}{4\pi}\int_0^{2\pi}\sqrt{(r_e + 2\rho\sin\phi)^2 + \rho^2\cos^2\phi}\, d\theta$$

Project $\bar{r}_e$ onto x-y plane,

$$\bar{r}_e = \frac{1}{4\pi}\int_0^{2\pi} r_e d\theta + \frac{1}{4\pi}\int_0^{2\pi}(r_e + 2\rho\sin\phi)d\theta$$

$$= \frac{1}{2}r_e + \frac{1}{2}r_e + \frac{1}{4\pi}\int_0^{2\pi} 2\rho\sin\phi d\theta$$

$$= r_e + \frac{1}{4\pi}\int_0^{\pi} 2\rho\sin\phi d\phi$$

$$= r_e(1 + \frac{\rho}{\pi r_e}), Let \frac{2\rho}{r_e} \equiv \alpha$$

$$= r_e\left[1 + \frac{1}{2}(\frac{\alpha}{\pi})\right]$$

The second term is identical to the result of the anomalous magnetic moment calculated from QED by taking into account of radiative correction. [61]

Equating the two expressions for $\alpha$ $\frac{2\rho}{r_e} = \alpha$ and $\alpha = \frac{r_0}{2r_e}$, one arrives at $r_0 = 4\rho$—the total width of the Hubius Helix. Now that $r_0$ has an explicit geometrical meaning, consequently the $\alpha$ can be interpreted geometrically as the ratio of the half (full) width to the radius (diameter) of the Hubius Helix. Thus, the present formulation has provided a geometric interpretation of the fine structure constant and the anomalous magnetic moment, bestows an explicit geometrical meaning on the classical radius of the electron, and justified the relation $\frac{e^2}{r_0} = mc^2$.

### 4.10. Charge as a Geometrical Attribute of the Hubius Helix

The charge as a physical attribute of an elementary particle concerns the following fundamental aspects. There are two types of charge in nature assigned positive and negative, and the charge attached to a particular elementary particle has a constant numerical value in a given system of units of physics. The effort to understand the definite value of the charge—charge quantization has followed at least two major routes, the magnetic monopole approach and flux quantization approach. [40, 62] The former approach, with the initial intention of looking for some explanation of $\frac{\hbar c}{e^2} \approx 137$, leads to a concept of magnetic monopole that sets up a quantization condition for the charge. The state of affairs is best described in Dirac's own words: "If there exists any monopole at all in the universe, all electric charges would have to be such that $e$ times this monopole strength is equal to $\frac{1}{2} n\hbar c$."[63]. The introduction of the monopole was not considered a complete solution of the problem of charge quantization, for it could not fix any value for $e$. [63] As for the latter approach it is not clear as to which of the two concepts, the charge quantization or the flux quantization, is the primary concept.

Now that $\alpha$ has found its geometrical meaning in subsection 4.9., through $\alpha = \frac{2\rho}{r_e} = \frac{r_0}{2r_e} = \frac{e^2}{\hbar c}$, the charge is likely to be accounted for geometrically as well. To render the charge a geometrical attribute, it is necessary to identify some geometric feature of the Hubius Helix with the charge. In order to do so, the Hubius Helix needs to be further examined in a deeper mathematical sense.

In geometry a space curve is characterized by three variables, the linking number $L_k$, the writhing number $W_r$, and the total twist number $T_n$, which are related through $W_r = L_k - T_n$. [64, 65] The $L_k$ is a topological property, whereas the $W_r$ and the $T_n$ are metrical properties. If a strip $S(X, U)$ is defined as a smooth curve $X$ together with a smoothly varying unit vector $U(t)$ perpendicular to X at each point, and $\omega_1$ is defined as the twist of the strip at each point of the curve $\omega_1 = \lim_{\delta s \to 0} \frac{\delta \phi}{\delta s}$, where the $\delta \phi$ is the angular displacement $\delta s$ apart along the strip; the *total twist* of the strip is defined as $T_w(X,U) = T_w \equiv \oint \omega_1 ds$, and the *total twist number* $T_n \equiv \frac{T_w}{2\pi}$. [65] (What applies to a strip does apply to an elastic rod or string, when the strip is considered as the generator of the latter. See also L. D. Landau and E. M. Lifshitz, Theory of Elasticity, Volume 7 of Course of theoretical Physics, Pergamon Press, London, 1959pp. 65-66, where the same quantity is called torsion angle.) Following Fuller's definition, F. Crick worked out a formula for calculating the total twist of a ribbon wound on a cylinder and going $N$ times, $T_w = N \sin \delta$, where $\delta$ is the pitch angle of the helix. [66] Applying this formula to the homotopy of the Hubius Helix employed in subsection 4.9., a circular-cross-sectioned wire being wound on a cylinder, going round two times in a right-handed helix, and joining the ends, we get $T_w = 2 \sin \delta$. For a small pitch angle, as in the present case, $\sin \delta = \tan \delta = \frac{\rho}{\pi r_e} = \frac{\alpha}{2\pi}$, thus $T_w = \frac{\alpha}{\pi} = \frac{e^2}{\pi \hbar c}$. $T_w$ is negative for the left-handed Hubius Helix. The disparity in the sign of the twist can therefore be identified with that in the sign of charge in the electron and the positron. Hence the twist reduced the charge—a physical attribute of the electron to a geometrical attribute of the Hubius Helix. At the same time, it bestows a geometric meaning for the coupling constant $\frac{\alpha}{\pi}$ in QED.

**4.11. Mass as a Geometrical Attribute of the Hubius Helix**

From the physical point of view, as a consequence of conjoining the special theory of relativity and quantum mechanics through the Hubius Helix the mass of an electron is related to the radius of the electron, $r_e = \frac{\hbar}{2mc}$. This relation resembles that of curvature and the radius of curvature of a plane curve, $\kappa = \frac{1}{r}$. The two relations become identical under the conditions $c = 1$ and $\frac{\hbar}{2} = 1$. The conditions are not at all unreasonable, but rooted on sound physical basis, the former as result of treating space and time on the same footing and the latter as result of choosing the electron's action as the basic unit of action.

Thus, from the mathematical point of view, under a suitable choice of the system of units, mass is analogous, if not identical, to curvature. The attempt to identify mass with curvature makes physical sense in two ways. For the first, the more massive the particle, the smaller the particle, and the more curved the space near the particle. For the second, a particle of zero mass like the photon travels in a straight line. Upon accepting this identification, the origin of mass can be understood as action in a curved space.

**4.12. Certain Transformation Properties of the Hubius Helix**

Physical laws expressed in terms of a system of equations of space time variables necessarily possess/lack certain transformation properties or symmetries, for instance being invariant/variant under translation or rotation with respect to a certain coordinate system. The general transformation properties of the Hubius Helix shall be worked out in a separate occasion by treating the Hubius Helix as a dynamic system, and examining such a system represented in Hamiltonian or Lagrangian form undergone proper Lorentz transformations (i.e., continuous Lorentz transformations that involve neither space nor time inversions). Before this is done, it is possible to work out the transformation properties of the Hubius Helix undergone three important discrete transformations, charge conjugation (C), space inversion or parity (P), and time inversion (T) and the combinations thereof to examine the relation between the electron and positron in particular, and matter antimatter in general.

The geometrical rules of correspondence of the electron to the positron formulated in subsection 4.6., the identification of the charge to the twist, and the connection of the sign of the twist to the parity of a Hubius Helix imply an association of charge with the parity (i.e., when the parity of the Hubius Helix changes, the sign of the twist is reversed, and vise versa). Thus C and P merge to an inseparable transformation (denoted as <u>CP</u> with the underline to emphasize the inseparability). However, the T transformation remains an independent one. The results of {<u>CP</u>, T, <u>CP</u>T, T<u>CP</u>} transformations can be envisaged on the electron and positron at their respective spin up and down states $(e_u^-, e_d^-, e_u^+, e_d^+)$. <u>CP</u> changes the electron/positron to the positron/electron, but keeps the spin orientation unchanged. T does not change the electron/positron to the positron/electron, but reverses the orientation of the spin. <u>CP</u>T changes the electron/positron to the positron/electron, as well as the orientation of the spin. <u>CP</u>T and T<u>CP</u> are commutative transformations. These results indicate that the matter antimatter worlds as represented by the electron (left-handed $L$) and the positron (right-handed $R$) are governed by rules $R \neq L, \overline{R} \neq \overline{L}$, but $\overline{R} = L, R = \overline{L}$.

**4.13. The Dimension of the Electron in Particle Physics Perspective**

It is a general belief that the electron is a point-like particle with almost no measurable dimensions, and that the electron does not possess any sub-constituent. This belief is largely based on the results of Møller and Bhabha scattering, obtained from many experiments performed in the last twenty years at PETRA (Positron-Electron Tandem Ring Accelerator Facility at DESY laboratory in Hamburg), PEP(The Positron-Electron Project, a collaborative effort of SLAC and Lawrence Berkeley Laboratory), TRISTAN (The $e^+e^-$ Collider at National Laboratory for High Energy Physics, KEK in Japan), and LEP (The Large Electron Positron Collider at CERN). In the experiments conducted at PEP in the tests of leptonic substructure, Bender et al. stated "experimentally there is no indication that the electron has structure [composite of more fundamental particles].

Lower limits for cutoff parameters are in the 100—200 GeV range, corresponding to an electron size of less than $10^{-16}$ cm. Therefore, if the electron is a composite particle its constituents are strongly bound, giving the electron the observed point-like quality at experimentally accessible energies." [67] As a matter of fact the results obtained from the scattering experiments have been a "no-go" sign to any attempt of the so called large-electron theories, where the size of the electron was in an order of its Compton wavelength. [11] The present work, where the electron is formulated as having a defined structure, characteristic dimensions, and no substructure, does not receive any exception—the dimension of the electron inferred from those experiments is several orders of magnitude smaller than the radius of the electron derived from the present work $r_e = \dfrac{\hbar}{2mc} = 1.93 \times 10^{-11} cm$.

This apparent discrepancy presents a grave challenge to the present work. How can such a large electron manifest such a small size in scattering? Or, why the scattering experiments reveal a point-like particle, if the size of the electron is in an order of its Compton wavelength? In order to answer these questions, let us examine what happens, when a particle is moving at high energy. First of all, when a particle is at rest, it possesses rest energy $E = mc^2$. When it is moving, its total energy becomes $E = mc^2 \left( \dfrac{1}{\sqrt{1-(\dfrac{v}{c})^2}} \right)$ or its mass becomes $m' = m \left( \dfrac{1}{\sqrt{1-(\dfrac{v}{c})^2}} \right)$, according to the relativistic dynamics. The change in mass may result in either "a point of view" that although the particle becomes heavier as it moves, it is still the same particle, or "a point of view" that since the particle is heavier as it moves, it has become another particle. Based on the first "point of view", the formula $r = \dfrac{\hbar}{2mc}$ derived in the present work suggests that for a given particle as the mass of the particle increases, the size of the particle decreases. Based on the second "point of view" the formula suggests that for different particles the heavier the particle, the smaller the dimension. Combining $E = mc^2$ and $r = \dfrac{\hbar}{2mc}$, we can obtain the radius of the electron as a function of its total energy $r = \dfrac{\hbar c}{2E} = \dfrac{1.97}{2E(GeV)} \times 10^{-14} GeVcm$. This relation demonstrates how a Compton-sized electron manifests a point-like particle in scattering experiments, and provides the reason why high energy scattering reveals the electron as a point-like particle. According to this relation, when an electron is accelerated to the energy of 100 GeV or 200 GeV, its diameter has diminished to $1.97 \times 10^{-16}$ cm or $0.99 \times 10^{-16}$ cm respectively. These values agree well with the dimension of the electron inferred from the high energy scattering experiments. [67]

## 5. REFURBISHMENT OF THE POSTULATE

From the original postulate put forward in section 3 that the topological structure of the electron is a closed two-turn helix (a so-called Hubius Helix) that is generated by circulatory motion of a mass-less particle at the speed of light, a system of parameterized equations emerged. With the help of the concepts in the classical physics, the equations tie together most, if not all, attributes of the electron, particularly the anomalous magnetic moment:

$$\begin{cases} x = (r_e + \frac{1}{2}r_0 \sin\omega_d t)\cos\omega_Z t \\ y = (r_e + \frac{1}{2}r_0 \sin\omega_d t)\sin\omega_Z t, \\ z = \frac{1}{2}r_0 \cos\omega_d t \end{cases}$$

where $r = \sqrt{x^2 + y^2 + z^2}$, $\bar{r} = \frac{1}{4\pi}\int_0^{4\pi} rd\theta$, $r_e = \frac{\hbar}{2mc}$, $\omega_d = \frac{mc^2}{\hbar}$, $\omega_Z = \frac{2mc^2}{\hbar}$,

$r_0 = 0, (0 \leq \omega_d t \leq 2\pi, 0 \leq \omega_Z t \leq \pi), r_0 = \frac{e^2}{mc^2}(2\pi \leq \omega_d t \leq 4\pi, \pi \leq \omega_Z t \leq 2\pi)$,

$\alpha = \frac{r_0}{2r_e}$, $T_w = \frac{r_0}{2\pi r_e} = \frac{\alpha}{\pi}$.

## 6. IMPLICATIONS

### 6.1. Issues Inherent in the Point-like Electron

That electrons are point-like particles has been a statement in perhaps all occasions of the description of the electrons as elementary or fundamental particles. However, to accept the statement as a concept, one must choose to live with certain issues inherent in the concept.

"The *self-energy* of the electron has been an embarrassment for almost a century, and in a sense for longer than there have been authentic electrons (first demonstrated by J. J. Thomson in 1901) and for far longer than physics has been complicated (but, some would say, simplified) by quantum mechanics. For if electrons are strictly point-like object, their self-energy, most simply taken as their potential energy in the electric field they carry, must be infinite. … People like H. A. Lorentz were well aware of the difficulty. … In itself, that would be no great scandal, even if electrons were point-like entities, for the absolute value of a potential energy is unimportant. But by 1905,

Einstein had sharpened the dilemma by demonstrating the equivalence of energy and mass. How, then to escape from the conclusion that a point-like electron should have *infinite mass*?" [68] (My *Italics*) The scope of the issues was expanded, when the spin and the magnetic moment were assigned to the electrons [38], for there is no room in a point-like electron to accommodate for such a large magnetic moment, and the *magnetic self-energy* associated with it.[69]

It is believed that a finite electron is the most natural concept that makes the total energy of the Coulomb field of the electron finite [24], therefore a way out of the dilemma.

Now that a finite structure has been postulated with the consequences being worked out, the dilemma should be resolved. Concerning the relation of the Coulomb self-energy and the rest energy, $\frac{e^2}{r_o} = mc^2$ was already justified in subsection 4.10. Concerning the relation of the Coulomb self-energy and the magnetic self-energy, it is rather straightforward to show that the electric field is indeed equal in magnitude to the magnetic field, $\frac{e}{r_e^2} = \frac{\mu_e}{r_e^3}$ at $r = r_e = \frac{\hbar}{2mc}$, taking $\mu_e = er_e$ from subsection 4.5. This relation was derived a long time ago, though the problem was formulated in reverse order. [34] The relations, $\frac{e^2}{r_0} = mc^2$ and $\frac{e}{r_e^2} = \frac{\mu_e}{r_e^3}$ should thus provide a basis for a self-consistent description of the finite self-energy to finally resolve the issue of infinite self-energy. The finite size through $r_e = \frac{\hbar}{2mc}$ has led to a finite rest mass. Furthermore, the relation of the radius to the relativistic mass as formulated in subsection 4.13 should lead to the account for the Lorentz contraction.

## 6.2. Other Kinds of Particles

One does not distinguish particles from antiparticles when counting the kinds of particles. Following this convention, one includes antiparticles when discussing the particles in general. In section 4, the physical attributes of the electron, including mass, charge and spin were formulated in terms of geometrical and topological attributes of the postulated structure, the Hubius Helix; for instance, the spin $\frac{1}{2}\hbar$ as the consequence of $4\pi$ periodicity, charge as the consequence of twist, and mass as the consequence of curvature. Thus, particles of the same geometrical/topological attributes may be classified as the same kind or put into the same geometrical/topological class. Fermions with $s=1/2$, $m \neq 0$, and charge $e$ may be considered as one class— Hubius Helices of the same twist but different radii, whereas massive bosons with $s=1$, $m \neq 0$, and charge=0 may be considered as another geometrical class—circles of different radii. In addition, photons with charge=0, $m=0$ can be considered as yet another class—strings of zero curvature. In this classification regime, a muon and an electron, which have identical properties except for mass and life time, belong to the same geometrical/topological class. The applicability of the classification regime is affirmed in the case of the electron and the muon by the fact that the formulation applied to the electron is equally applicable to the muon before it decays, provided that $\alpha$ is kept the same for the muon as for the electron. However, for protons the compound nature and large anomalous magnetic moment seem to suggest a quite different class of geometric structure.

## 6.3. The Physical Constants and Physical Theories, Fundamental Length, and the Constants of Nature

Physics with experiment and theory as its fundamental components deals with physical quantities in practice. A physical quantity is said to be a physical constant, if it is time invariant, in spite of its varied numerical value in different systems of units. A physical constant is further said to be fundamental or universal, if it is universally invariant, that is if it is not too specific, too closely associated with the particular properties of the material or system upon which the measurements are carried out. The Fundamental Physical Constants compiled by CODATA appear to bear such quality. [70] The importance of the fundamental constants lies in the fact that they, through association with physical theories, constitute to a varied extent a unified description of nature or natural phenomena.

However, their association with the respective physical theories does not seem to indicate that they are all equally fundamental. Furthermore, the interplay between experimental and theoretical physics on the physical quantities does indicate that they are inter-expressible. These led to a speculation fifty years ago that in a complete physical theory all the constants would be expressible in terms of only four basic constants: the Boltzmann's constant $k_B$, the speed of light $c$ (the Einstein's constant, following a recent proposal of Kenneth Brecher of Boston University), the Planck's constant $h$, and the elementary length $\lambda$, "the quadrumvirate of complete physical theory" according to Gamow, [71] or "universal constants of the first kind" according to Heisenberg. [55] While $k_B$ is responsible for the connection between isolated entities and their ensemble, the rest are concerned with the so called microphysics. Being 'universal constants of the first kind', $h$ and $c$ share the virtue that they emerge as a consequence of conjoining physical theories, they are of pivotal importance in the new theories, and they designate the limits in whose proximity our intuitive concepts can no longer be used without misgivings. $c$ is a consequence of conjoining Newtonian mechanics and the Maxwell theory, reflects the relation between the time-component and space-components of the four-Minkowski space, appears as a constant in the basic equation $E = mc^2$ on which the special theory of relativity rests, and designates the highest possible value of any physical velocity. $h$ is a consequence of conjoining the statistical theory of heat and the Maxwell theory, appears as a constant in the basic equation $E = h\nu$ on which quantum mechanics rests, and designates the smallest possible value of action concerning absorption-radiation process. Unlike $h$ and $c$, the fundamental length does not have an association with existing fundamental theories as such, and the argument for its inclusion in the complete physical theory was mainly based on dimension analysis and the fact that a length is a basic measure in physics. [55, 71] The argument is as follows: there must exist [besides $h$ and $c$] another 'universal constant of the first kind' having the dimension of a length or a mass, for a length or mass cannot be formed dimensionally from $h$ and $c$. And the physical significance of this constant is expressed more clearly by introducing it as a fundamental length.

The introduction of the fundamental length to physics in some natural way was considered one of the problems in the post-relativity and quantum mechanics physics. [72] And the discussion of questions associated with it appears to be the most urgent task. [55] But eventually "How this new constant of elementary length can be introduced into the existing formalism of physical theory is, of course, anybody's guess." [71] In analyzing the relation between the fundamental constants and physical theories, it was envisioned that "when quantum theory and relativity wave theory are conjoined, account must be taken of a universal constant having the dimension of a length." [55] In other words, the emergence of the fundamental length ought to be a consequence of conjoining quantum theory and relativity theory.

The identification of the radius of the electron $r_e$ with the fundamental length $\lambda$ seems to be a natural way to introduce the fundamental length to physics, and a fulfillment of Heisenberg's vision and Gamow's speculation; for $r_e$ emerges as a consequence of conjoining quantum mechanics and relativity theory. It is of pivotal importance in the whole formulation as seen through the relation between the mass and the radius of the electron and the relation between the radius and the classical radius of the electron—the geometrical account for $\alpha$ and the electronic charge. Furthermore, it, together with $r_0$ designates the geometrical constraints that underlie the physical attributes of the electron. The fundamental length of other elementary particles is likely to be the same as the mass—all possess the same measure, but each has its own value. Thus, the fundamental length, as a basic measure of the elementary constituent of matter, deserves a position of being 'universal constant of the first kind' just like $h$ and $c$ in relation to physical theories, or a position in Gamow's quadrumvirate. I would vote Boltzmann's constant $k_B$, Einstein's constant $c$, Planck's constant $h$, and the radius of the electron $r_e$, (why not call it Hubius constant?) for the constants of nature.

## 7. CONCLUDING REMARKS

The fundamental ways of describing phenomena in nature are based on the concepts of particle, field, wave, and string [‡]. These concepts foster powerful physical theories on the one hand, and constitute different philosophical views of the physical universe on the other. The concept of massive point particles— being introduced as an abstraction of macroscopic objects in mechanical motion where the internal structure of the objects can be totally ignored—evolved into what is called the particle view, fostered Newtonian mechanics, Maxwell-Boltzmann's theory of heat, and led ultimately to the concept of elementary particles. The concept of lines of force—being conceived to visualize the medium, through which electromagnetic forces propagate—evolved into the field view, fostered the theories of electromagnetism of Faraday, Maxwell and Hertz and Einstein's field theory of gravitation, and led ultimately to the concept of unified field theories. The concept of wave-particle duality—being derived from the undulation and corpuscular nature of light and matter—evolved into the wave-particle or quantum field view, fostered the wave mechanics and quantum field theories, and led to the concept of the grand unification—the unified quantum field theory of elementary particles and interactions.

While the particles, fields, and waves at the present stage of their development are truly abstract representation in nature; the Hubius Helix postulated above portrays an intuitive picture for the electron. As demonstrated in the previous sections, the Helix allows us to apply classical concepts to it. With help of these concepts, it reproduced the static and dynamic spectroscopic properties of the electron. The very virtue of the Helix lies in the fact that a single piece of mathematical fact is able to tie together all of the attributes of the electron identified in Section 1.

With respect to the relation to the philosophical views, the Helix defies the particle view, so far as the electron is concerned. It is consistent with the wave-particle view, for the formulation and the wave mechanics share the same starting point, the same standing wave argument for quantization, and obtained the same wavelength-momentum relation for the electron. As the present work limited itself within the context of the electron as an isolated entity, the discussion of possible impact of the present work on the fundamental postulates of quantum mechanics is not intended. As for the field view, in spite of the relation between the Hubius Helix and the Möbius strip or non-trivial fiber-bundles that are closely connected to gauge field, [73] the formulation of the mathematical relation between the Hubius Helix and the gauge field is not attempted.

To conclude, the present work—as the result of an endeavour of a historical, physical, mathematical, and philosophical study of the electron as concept, isolated entity and theories—is a manifest of my belief in Einstein's conviction "You know, it would be sufficient to really understand the electron." [74]

**Footnotes:**

† This pullet is different from the rest ones in that it does not imply the Zitterbewegung as a well established concept or basis of the argument in the present paper rather as a mathematical consequence of the Dirac equation derived from a particular viewing angle, whose physical existence and interpretation are debatable even to the present day. And our understanding of it is still far from complete.

¶ The formulation does not treat the electron in conventional mechanical terms like in Newtonian mechanics, where all forces exerted on the object concerned are identified and put into the Newton equation in order to find the equation of motion; nor like in Hamilton mechanics or Lagrange mechanics, where the dynamical variables are defined, and the Hamiltonian or Lagrangian of the system is put into the respective equations in order to find the equation of motion. Instead, the formulation treats the electron as a space time structure by accepting the attributes of the electron (enlisted in section 1) as facts and correlating them to the geometrical or topological attributes of the space time structure. As consequence of such an approach, the framework of the paper, even if it can be tentatively defined as the space time structure, seems somewhat ambiguous, and some equations cited appear out of context. Since one of the expectations of the paper is to provide a foundation on which improved understanding of the electron can be built, the acceptance of the present framework or the emergence of a new framework and proper context is left to the future development on the basis of this paper.

‡ The relevance of the string to the Hubius Helix seems somewhat questionable, therefore the relation between the string and the Hubius Helix, and between the string theories and the present work will not be discussed.